\documentclass{article}

\usepackage[preprint]{neurips_2023}

\usepackage[utf8]{inputenc} 
\usepackage[T1]{fontenc}    
\usepackage{url}            
\usepackage{booktabs}       
\usepackage{amsfonts}       
\usepackage{nicefrac}       
\usepackage{microtype}      
\usepackage{xcolor}         
\usepackage{graphicx}
\usepackage{amsmath,amssymb}
\usepackage{physics}
\usepackage{cleveref}
\usepackage{siunitx}

\title{Environmental effects on emergent strategy in micro-scale multi-agent reinforcement learning}

\author{%
  Samuel Tovey \\
  Institute for Computational Physics\\
  University of Stuttgart\\
  Stuttgart, Germany 70569 \\
  \texttt{stovey@icp.uni-stuttgart.de} \\
\And
  David Zimmer \\
  Institute for Computational Physics\\
  University of Stuttgart\\
  Stuttgart, Germany 70569 \\
  \texttt{dzimmer@icp.uni-stuttgart.de} \\
\And
  Christoph Lohrmann \\
  Institute for Computational Physics\\
  University of Stuttgart\\
  Stuttgart, Germany 70569 \\
  \texttt{clorhmann@icp.uni-stuttgart.de} \\
\And
  Tobias Merkt \\
  Institute for Computational Physics\\
  University of Stuttgart\\
  Stuttgart, Germany 70569 \\
  \texttt{tmerkt@icp.uni-stuttgart.de} \\
\And
  Simon Koppenh\"{o}fer \\
  Institute for Computational Physics\\
  University of Stuttgart\\
  Stuttgart, Germany 70569 \\
  \texttt{skoppenheofer@icp.uni-stuttgart.de} \\
\And
  Veit-Lorenz Heuthe \\
  Department of Physics - Soft Condensed Matter\\
  Universit\"{a}t Konstanz\\
  Konstanz, Germany \\
\And
  Clemens Bechinger \\
  Department of Physics - Soft Condensed Matter\\
  Universit\"{a}t Konstanz\\
  Konstanz, Germany \\
\And
  Christian Holm \\
  Institute for Computational Physics\\
  University of Stuttgart\\
  Stuttgart, Germany 70569 \\
  \texttt{holm@icp.uni-stuttgart.de} \\
}

\begin{document}

\maketitle

\begin{abstract}
    Multi-Agent Reinforcement Learning (MARL) is a promising candidate for realizing efficient control of microscopic particles, of which micro-robots are a subset.
    However, the microscopic particles' environment presents unique challenges, such as Brownian motion at sufficiently small length-scales.
    In this work, we explore the role of temperature in the emergence and efficacy of strategies in MARL systems using particle-based Langevin molecular dynamics simulations as a realistic representation of micro-scale environments.
    To this end, we perform experiments on two different multi-agent tasks in microscopic environments at different temperatures, detecting the source of a concentration gradient and rotation of a rod.
    We find that at higher temperatures, the RL agents identify new strategies for achieving these tasks, highlighting the importance of understanding this regime and providing insight into optimal training strategies for bridging the generalization gap between simulation and reality.
    We also introduce a novel Python package for studying microscopic agents using reinforcement learning (RL) to accompany our results.
\end{abstract}

\section{Introduction}
In recent years, machine learning (ML) and physics have enjoyed a strong relationship in broad fields ranging from quantum chemistry~\citep{zaverkin23a} to cosmology~\citep{krippendorf22a}.
An emerging application of ML in the physical sciences is understanding the dynamics and collective behavior of microscopic particles or colloids~\citep{zoettl}.
In these low Reynolds number regimes, the physics of Brownian motion and hydrodynamics dominate, leading to exciting challenges~\citep{brewer22a}.
Despite these challenges, unusual collective behavior has been observed in both biological~\citep{takuma21a} and artificial~\citep{zoettl16a} active matter environments.
A current research area in this field is understanding how to program these artificial microscopic colloids or agents to achieve tasks like locating positions in an environment or moving objects. 
Such detailed control over microscopic agents can revolutionize many fields, with medicine being of great interest~\citep{soto20a}.
While there has been a significant amount of progress utilizing classical algorithms~\citep{moreau21a, lolli22a}, recently, reinforcement learning (RL) has become a promising candidate~\citep{qin23a, hartl21a, borra22a, landin21a}.
Most RL approaches thus far have centered around using Q-learning for single-agent tasks.
However, effective deployment of these agents will likely involve collective behavior and, therefore, a multi-agent setting.
Furthermore, the role of the environment in these tasks has yet to be investigated to a great extent, partially due to many groups relying on experimental implementations of these agents and, therefore, not having access to such variables.
This also prohibits the training time of the models, as experimentally driven RL is laborious. 
In this work, we approach the problem of understanding how the Brownian motion of these agents impacts their emergent strategy and its efficacy at different temperatures.
We utilize a powerful physics engine designed to study these systems, ESPResSo~\citep{weik19a}, to perform these experiments at scale.
Experiments involve implementing MARL for the problems of detecting target points in a box and rotating a rod.

Our main contributions are:
\begin{itemize}
    \item We demonstrate emergent collective behavior in RL-driven micro-robots in realistic simulations.
    \item We discuss the role of Brownian motion in the agents' success in their tasks.
    \item We demonstrate the robustness of MARL algorithms against environmental noise.
    \item We highlight the differences in emergent strategies as a function of temperature and comment on this as a necessary consideration in bridging the generalization gap with experiments.
\end{itemize}

All of the infrastructure for the work performed in this investigation is packaged into an open-source Python project, SwarmRL, which is available publicly on GitHub.

\section{Related Work}
Much work has been dedicated to both emergent strategy in microscopic agents and the application of RL at these length scales.
A lot of this work has focused on understanding relationships between artificial microswimmers and their biological counterparts or applying simplified RL strategies such as Q-learning to various problems.
\paragraph{Collective Behaviour in Artificial Microswimmers}
Understanding the behavior of organisms on microscopic scales has been a central research focus since the early 1900s.
In recent years, with the invention of artificial microswimmers, improved imaging methods, and advanced computational systems, one has been able to probe these behaviors in much greater detail.
Of particular interest is the emergence of collective behavior in groups of microswimmers; that is, with limited local information about their environments, groups of agents can share knowledge to achieve a task.
In~\citet{lavergne19a} and~\citet{baeuerle20a}, group formation and behavior were studied utilizing classical interaction algorithms for microswimmers under varying conditions.
These studies demonstrated the emergence of collective behavior for swimmers with varying degrees of sensory ability.
In their 2022 paper,~\citet{chen22a} investigated the response of microscopic swarms to external threats.
This work emphasized that collective groups of microswimmers use information sharing to respond to events that individuals may not be aware of.
In each of these studies, while the effects of Brownian motion are implicit in the setup, the role of these fluctuations is not explored, nor is a specific task programmed.
\paragraph{Microscopic RL}
In the direction of RL, several studies have approached the problem of organizing microswimmers to the end of achieving a goal.
These studies typically rely on RL as a policy development algorithm.
\citet{qin23a} utilize a Q-learning algorithm to learn the swimming strategies of microswimmers and identify new swimming gaits for linked artificial micro-robots.
In the direction of learning policy,~\citet{hartl21a} have used a genetic algorithm to reproduce biological chemo-taxis in artificial swimmers and~\citet{borra22a} implement actor-critic RL on a predator-prey problem and found that agents could utilize hydrodynamic cues to avoid predators in their environment.
Finally,~\citet{landin21a} utilized a Q-learning approach to study how the environments of microswimmers impact their learning process for individual swimmer tasks, including the role of Brownian motion.
However, they did not investigate the impact on collective behavior or for more complex tasks which require algorithms beyond Q-learning.

To date, the role of the stochastic forces experienced by microswimmers in their emergent strategy has yet to be investigated.
However, these forces are a crucial component of micro-robotics, and therefore, their impact on learning is identified here as the primary research gap we aim to fill.

\section{Problem Description}
\label{sec:problem-description}
Controlling multi-agent systems is a challenging task in-and-of itself. 
Additionally, on microscopic agents' length scales, thermal fluctuations are relevant.
To varying degrees they will prohibit agents from performing a task by introducing a random component preventing the outcome of an action from being predicted with certainty.
On the contrary, the existence of this Brownian motion will encourage exploration of the agents as each action they take is perturbed.

We study numerically a system comprising colloids on the micrometer scale that can perform actions such as actively moving or rotating.
Using RL, we want to enable the colloids to solve tasks like finding the source of an attractant or rotating a rod.
We set out to understand the role of thermal fluctuations in the efficacy of RL on microscopic scales and the emergent strategy adopted by the agents.

In all simulations, a Langevin thermostat correctly reproduces the agents' Brownian motion and maintains the system's temperature in a statistical physics sense.
More broadly, stochastic effects can also occur at length scales where thermal motion is not relevant, for example imperfect motors or sensors on a robot.
Our work is therefore also relevant for applications beyond the micro-scale.

\section{Methods}
The successful implementation of these experiments has required many unique components.
Here we outline briefly the critical aspects of the work.

\subsection{Particle Simulations}
We simulate the trajectories of particles using the overdamped Langevin equations of motion for position and orientation
\begin{equation}
      \dot{\vb{r}}_i = \frac{1}{\gamma_t} \left[F(t) \vb{e}_i(\Theta_i) - \nabla V(\vb{r}_i, \{\vb{r}_j\}) \right] + \sqrt{2 k_B T / \gamma_t} \vb{R}^t_i(t), 
      \label{eq:brownian_pos}
\end{equation}
\begin{equation}
      \dot{\Theta} = \frac{1}{\gamma_r} \tau(t) + \sqrt{2 k_B T / \gamma_r} R^r_i(t).
      \label{eq:brownian_angle}
\end{equation}
Here, $\vb{r}_i$ is the (two-dimensional) position of particle $i$, $\Theta_i$ the angle describing the particle orientation,  $\gamma_{(t,r)}$ the translational (rotational) friction coefficient, $F$ and $\tau$ an active force and torque corresponding to an action, $\vb{e} = (\cos(\Theta), \sin(\Theta))^T$ the particle orientation, $V$ an interaction potential between all particles in the system, $k_B$ the Boltzmann constant, $T$ the temperature and $\vb{R}^{(t, r)}_i$ a noise term with zero mean and correlations according to $\expval{R^{(t,r)}_i(t) R^{(t,r)}_j(t')} = \delta_{ij} \delta(t-t')$, where $\expval{\cdot}$ denotes an ensemble average.
For colloids of radius $a$ in a fluid with dynamic viscosity $\mu$ we calculate the friction coefficient according to Stokes' law as $\gamma_t = 6 \pi \mu a$ and $\gamma_r = 8 \pi \mu a^3$. 

\Cref{eq:brownian_pos,eq:brownian_angle} are solved numerically using the ESPResSo~\citep{weik19a} simulation package with a time-step $\delta t = \SI{0.01}{\second}$, the actions that determine $F(t)$ and $\tau(t)$ are updated every time slice $\Delta t = \SI{1}{\second}$.
In all cases, unless otherwise specified, when referring to time in this investigation, we refer to the time slice, i.e., the number of times an action is computed for each agent in the simulation.

We model particle interactions with the two-body Weeks-Chandler-Anderson (WCA)  potential~\citep{weeks71a}, which can be seen as an almost-hard-sphere interaction:
\begin{equation}
    V = 4\cdot V_{0}\left( \left(\frac{\sigma}{r_{ij}}\right)^{12} - \left(\frac{\sigma}{r_{ij}}\right)^{6} \right) + V_{0}.
\end{equation}
Here, $r_{ij} = || \vb{r}_i - \vb{r}_j ||_2$ is the absolute distance between the particles, and $\sigma = 2a$ the colloid diameter. We choose the interaction strength $V_0 = k_B T$.

\subsection{RL Architecture}
\begin{figure}[t]
    \centering
    \includegraphics[width=0.7\linewidth]{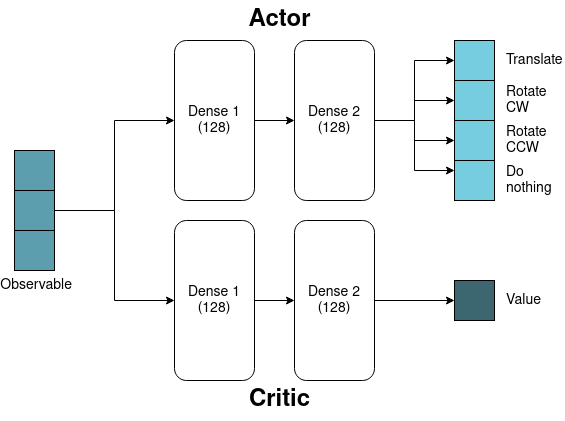}
    \caption{The workflow of the MARL utilized in the experiments. An observable is passed to an actor and a critic, each composed of two dense layers with 128 units. The actor returns a probability of selecting each of the four actions and the critic returns the estimated value of the state.}
    \label{fig:actor-critic}
\end{figure}

RL is a branch of machine learning in which an agent in state $s_t$ interacts with its environment by taking action $a_t$ that brings it to a following state $s_{t+1}$. 
The agent aims to maximize each state's scalar reward $r_t$. It does so by optimizing the policy $\pi$ that governs the agent's behavior. 
MARL as a subfield of RL is closely related to game theory and focuses on multiple such learning agents.
In addition to interacting with the environment, agents also interact with each other in cooperative or competitive ways~\citep{littman94a}. 
At every time slice of the experiment, each agent takes an action based on its observation. 
Each possible action at that time is sampled with a probability determined by the policy.
The Gumble-max trick~\citep{huijben22a} is used to sample this categorical distribution efficiently. 
The policy is updated using a policy gradient method in which the actor takes the role of the policy and the critic evaluates the actor's performance after each episode~\citep{sutton98a} . 
The actor and the critic are neural networks, each consisting of two dense layers with 128 neurons. 
The actor-critic architecture is shown in Figure~\ref{fig:actor-critic}.
The Adam optimizer with a learning rate of $\lambda = 0.001$~\citep{kingma17a} is used for the network parameter update. 

\subsection{Active Brownian Particles as Agents}
\label{subsec:janus-particles-as-agents}
\begin{figure}[t]
    \centering
    \includegraphics[width=.8\linewidth]{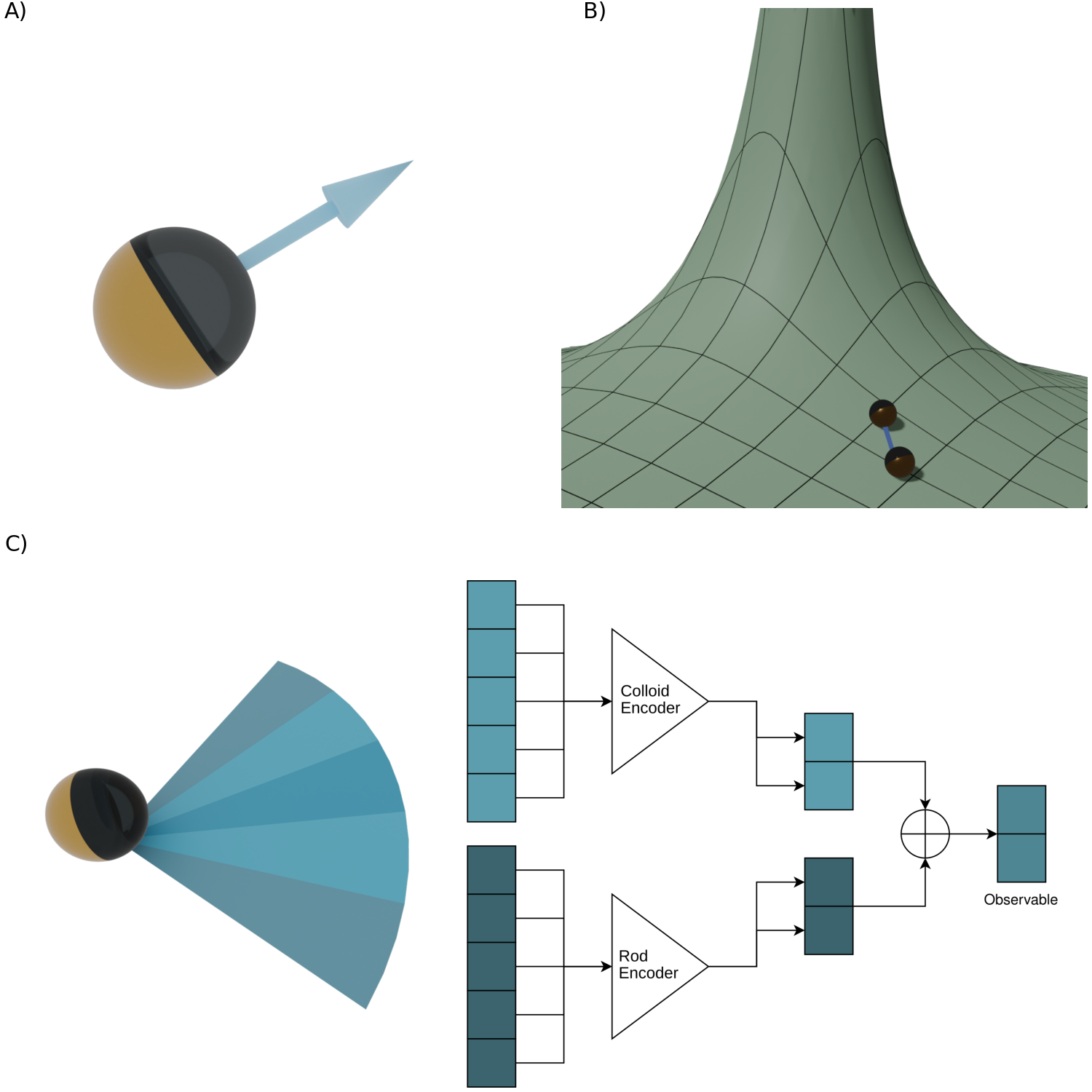}
    \caption{Various aspects of the agents deployed in this investigation. A) A representation of an ABP particle coated on one side. B) Visualization of the concentration sensing observable. C) Visualization of the vision cone observable including the trainable embedding layers.}
    \label{fig:rl-agents}
\end{figure}

We have used microscopic agents in all the experiments performed that mimic active Brownian particles such as the Janus particle~\citep{walther13a} shown in Figure~\ref{fig:rl-agents}A).
These particles can mimic active biological matter upon excitation from an outside driving force.
In experimental setups, this driving force is often one or more lasers applied along a specific axis of the colloid to induce either rotation or translation.
This axis is designated the \textit{director} of the colloid and emulates a forward-facing agent.
The driving mechanism is accurate enough to allow for a well-defined action space for each colloid in the system:
\begin{equation}
\mathcal{A}=
    \begin{cases}
        \text{Translate:} & F = 10.0\text{, } \tau = (0.0, 0.0, 0.0) \\
        \text{Rotate CCW:} & F = 0.0 \text{, }\tau = (0.0, 0.0, 10.0) \\
        \text{Rotate CW:} & F = 0.0 \text{, } \tau = (0.0, 0.0, -10.0) \\
        \text{Do Nothing:} & F = 0.0 \text{, } \tau = (0.0, 0.0, 0.0)
    \end{cases}
\end{equation}
where $F$ is the force magnitude in simulation units applied along the forward direction of the colloid and $\tau$ is the torque.
Another key feature of the reinforcement process is the state description parsed as an input for each agent.
In this investigation, two different state descriptions, referred to hereafter as observables, are utilized.
\paragraph{Concentration Sensing}
The most simple observable used in the experiments is termed concentration sensing.
This observable aims to replicate the sensing capabilities of biological organisms in their ability to register a change in an environmental factor such as temperature or a food source.
The observable itself computes a change in the concentration of some field defined in this study as a potential decaying like $1/r$.
At each action update, the models receive a value computed by:
\begin{equation}
    \centering
    o_{i}(t) = f\left(||\hat{\mathbf{r}}_{i}(t) - \hat{\mathbf{r}}_{s}(t)||_{2}\right) - f\left(||\hat{\mathbf{r}}_{i}(t-\Delta t) - \hat{\mathbf{r}}_{s}(t-\Delta t)||_{2}\right),
    \label{eqn:concentration-sensing}
\end{equation}
where $o_{i}$ is the observable for the $i^\text{th}$ colloid, $f$ is the chosen field, $\hat{\mathbf{r}}_{i}(t)$ is the position of the $i^{\text{th}}$ colloid at time $t$, $\Delta t$ is the amount of time since the last action was computed, $\hat{\mathbf{r}}_{s}(t)$ is the position of the source of the potential at time $t$, and $||\cdot||_{2}$ denotes a Euclidean norm.
In our experiments, the source remains fixed in space and will not vary in time.
The concentration sensing observable is shown graphically in Figure~\ref{fig:rl-agents}B).
\paragraph{Vision Cones}
The second observable employed in our studies was the vision cone shown graphically in Figure~\ref{fig:rl-agents}C).
This approach mimics an agent with the ability to identify particle species and a blurred sense of direction, that is, a range of angles within which it knows other particles are present.
In this study, the vision cone is broken into five components.
Each component of the vision cones computes $N$ numbers where $N$ is the number of unique species in the system, e.g., for rod rotation, there are active particles and rod particles resulting in $N=2$.
The value of the cone is computed by:
\begin{equation}
    \centering
    o^{jk}_{i} = \sum\limits_{n \in \mathcal{C}_{j}} \frac{1}{||\mathbf{r}_{i} - \mathbf{r}^{k}_{n}||_{2}},
\end{equation}
where $o^{jk}_{i}$ is the observable of the $i^{\text{th}}$ particle for the $j^{\text{th}}$ vision cone computed for all particles of species ${k}$ which lie within the cone represented by $\mathcal{C}_{j}$.
In order to provide this information to an actor and critic, we implemented an embedding scheme.
In this scheme, the vision cone values associated with the rods and colloids are split into two vectors and passed through their trainable embedding layers, producing a reduced, 2-dimensional output.
Element-wise addition of these vectors forms the final observable used to compute actions.
The embedding layers are trained along with the actor and critic during the simulations, allowing each model to learn a representation of its environment.

\subsection{SwarmRL}
\label{subsec:swarmrl}
\begin{figure}[t]
    \centering
    \includegraphics[width=0.9\linewidth]{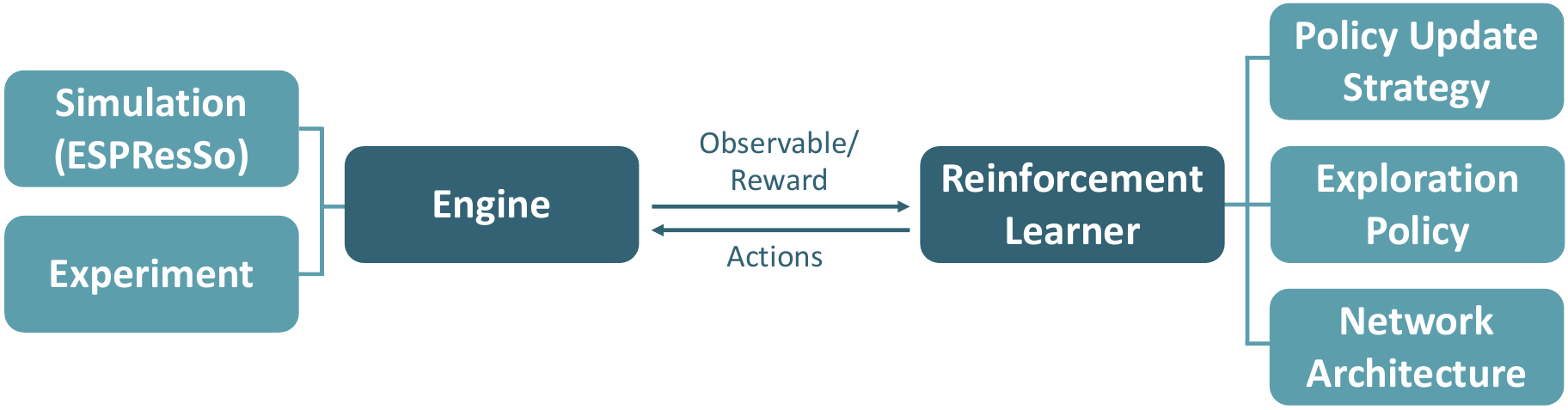}
    \caption{Simplified architecture of SwarmRL displaying data communication between the RL models and the experiments.}
    \label{fig:swarmrl-arch}
\end{figure}

The RL performed in this investigation is part of a project involved with not only training micro-swimmers but also deploying trained models in natural experiments.
Therefore, all of the infrastructure surrounding the training of the agents and deployment in the ESPReSso simulation engine has been written into an open-source Python package SwarmRL.
As illustrated in Figure~\ref{fig:swarmrl-arch}, SwarmRL provides an interface to each component of the RL pipeline as well as different \textit{environments} including both the simulation engine and actual experiment setups. 
SwarmRL is built on the JAX~\citep{bradbury18a} ecosystem and utilizes Flax~\citep{heek23a} for neural networks.
SwarmRL is publicly available on GitHub at~\url{https://github.com/SwarmRL/SwarmRL}.

\section{Experiments}
This investigation performs two experiments: source detection and rod rotation.
Each task requires a differing degree of complexity, and in each case, the agents discover different strategies depending on their environment.
Training for each task occurs in the same way.
Each model is trained with ten agents for a total of 10.000 episodes.
The first 5000 episodes are performed using an exploration rate of 20 \%; that is, 20 \% of the time, colloids chose a random action other than that provided by the actor.
The second 5000 episodes were performed without this exploration policy.
This is performed for ten ensembles with different starting simulation conditions and neural network initialization.
In all computations, an average value over these ensembles is computed, resulting in the corresponding error values.
At the end of the training, all the trained models are used in production simulations of 5.000.000 time-steps.
This procedure was carried out for five different temperatures: \{0 K, 150 K, 273 K, 300 K, 350 K \} to explore the changes in emergent strategy and efficacy of chosen strategy.

\subsection{Source Detection}
\begin{figure}[t]
    \centering
    \includegraphics[width=\linewidth]{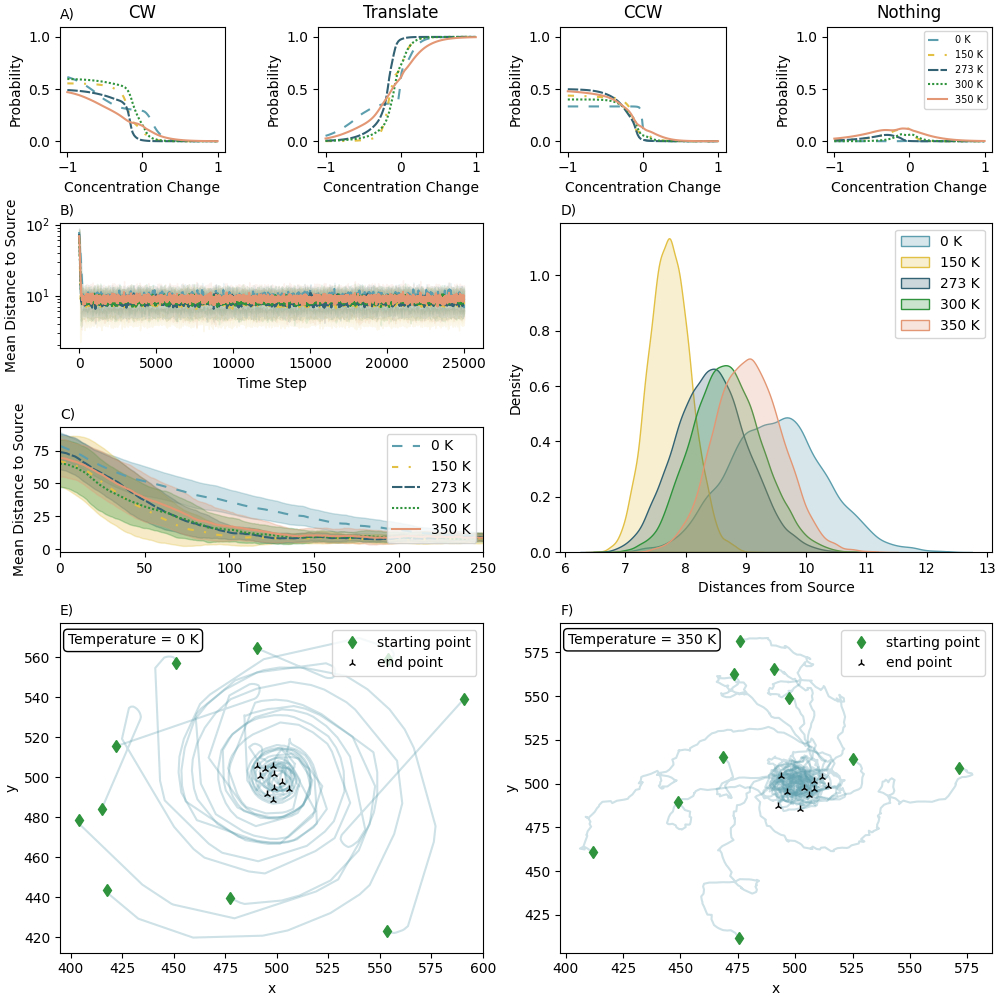}
    \caption{A) The policy evaluation performed on trained actors. Each figure shows the probability of taking a certain action given the input to the network. B) The mean distance of all agents to the source in the simulation as a function of time. Different line styles and colors correspond to different temperatures. C) A zoomed-in frame of B) where the higher temperature models converge faster and closer to the source of their sensing field. D) Histograms computed on the converged regions of the trajectory. E)-F) Example trajectory of the agents at 0 K and 350 K respectively.}
    \label{fig:find-location}
\end{figure}

The simplest task tested was that of source detection.
In this case, the colloids have a biologically inspired sense of smell; they can sense changes in some applied concentration field.
Their reward is based on changes in this field, i.e., moving closer to the source of the concentration yields a better reward.
Such a task is reminiscent of chemo-taxis in bacteria~\citep{zhuang16a} or, theoretically, detection of a drug delivery site in medicinal micro-robotics~\citep{schmidt20a}.
Mathematically, the reward closely resembles the observables of these colloids:
\begin{equation}
    \centering
    r_{i} = \alpha \cdot \text{Clip}\left(\frac{1}{||\hat{\mathbf{r}}_{i}(t) - \hat{\mathbf{r}}_{s}(t)||_{2}} - \frac{1}{||\hat{\mathbf{r}}_{i}(t-1) - \hat{\mathbf{r}}_{s}(t-1)||_{2}} \text{, } 0 \text{, None}\right)
\end{equation}
where $\alpha$ is some reward scaling value, the clip operation ensures that particles are rewarded for moving toward the source and receive no input if they move away.
The experiment results are outlined in Figure~\ref{fig:find-location}.
Figure~\ref{fig:find-location}A) illustrates the policy breakdown for inputs to the neural network at different temperatures.
To construct these plots, we pass artificial data through the trained models and compute the probability of different actions for these inputs.
We see that negative inputs in each temperature, i.e., a movement towards lower values of the concentration field, result in increased rotation probability and decreased translation.
This is opposed to movements towards the source, which increases translation probability.
Such a learned policy is reminiscent of run-and-tumble motion in bacteria in response to changing environments~\citep{ginkel21a}.
It appears that at all temperatures, the \textit{Do Nothing} actions is never included in the learned strategy and remains a randomly selected action.
We also noticed in these experiments that once the agents learned to rotate either clockwise or counterclockwise, they did not use the other action.

Figure~\ref{fig:find-location}B), C), and D) illustrate the distance of the colloids from the source at different temperatures.
In Figure~\ref{fig:find-location}C), the colloids trained at higher temperatures converge faster onto the source. 
This is expanded upon in~\ref{fig:find-location}D) where the histograms of the equilibrium region are plotted for each temperature.
These histograms show that models trained at higher temperatures converge closer to the source of the field and, in general, experience less variance around it.
This is generally true; however, it can also be seen that the 150 K model reached the closest to the source while higher temperatures moved back away. 
The mechanism behind this improved policy can be understood by looking directly at the trajectories of the models in Figures~\ref{fig:find-location}E) and F).
In the 0K simulation shown in~\ref{fig:find-location}E), the colloids undertake an orbital movement towards the source and, in doing so, maximize their reward over time.
In the case of a 350 K environment shown in~\ref{fig:find-location}F), the colloids cannot afford to take such a policy as the random fluctuations will push them out of their orbit.
In these cases, the colloids move more directly toward the source and perform more oscillatory movement once they are close.
This approach explains the improved proximity of the higher temperature models as they orient themselves to move directly to their target.
However, the stronger the Brownian forces acting on the colloids, the harder it will be to remain close to the source and the more significant are their fluctuations around it.
In summary, including Brownian forces in the model training results in a more direct approach to the source of the field attracting the colloids.
This policy results in closer average distances to the source and less variation around the it.

\subsection{Rod Rotation}
\begin{figure}
    \centering
    \includegraphics[width=\linewidth]{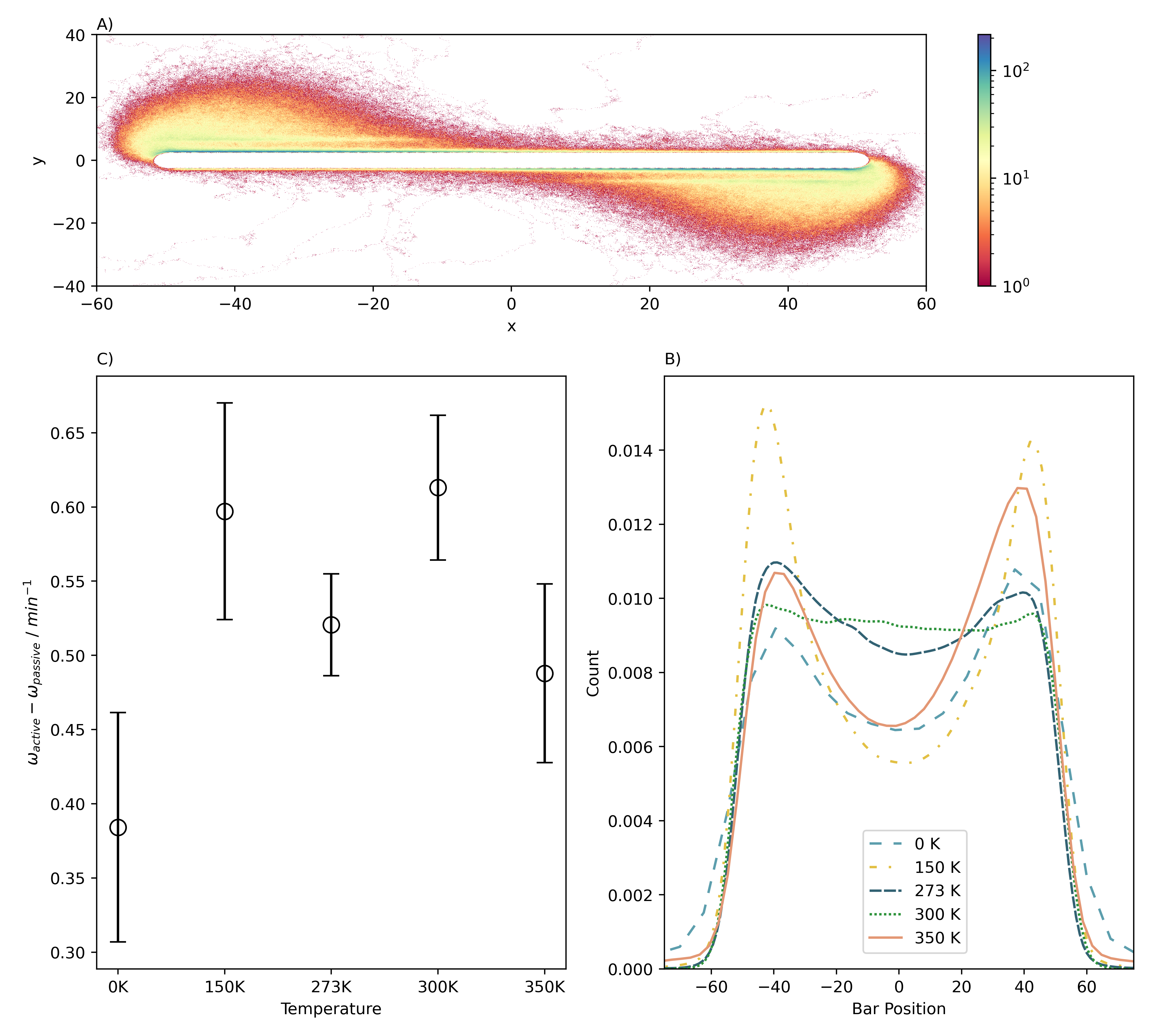}
    \caption{A) The 2D histogram formed using the trajectory of a single 150 K deployment simulation. B) The rod's angular velocity minus the corresponding value for a system without agents as a function of temperature. C) A histogram along the rod's long axis demonstrates the accumulation at the ends of the rod for each temperature.}
    \label{fig:experiment-2}
\end{figure}

Increasing the complexity of our experiments, the subsequent investigation involved training the agents to rotate a rod.
In these experiments, the rod is modeled as a series of rigidly bonded small colloids so that they cannot move away from one another.
This task is interesting as the colloids should cooperate to achieve this task efficiently.
The observables we use in the rod rotation experiments are the vision cones discussed in Section~\ref{subsec:janus-particles-as-agents}, which return values for both the agents and the rod colloids.
The rewards for this task are broken into two components, detecting the rod and increasing its angular velocity.
Rod detection is encouraged by rewarding the agents if they move closer to the rod particles:
\begin{equation}
    \centering
    r^{\text{rod proximity}}_{i} = \sum\limits_{j}\frac{1}{\hat{\mathbf{r}}_{i}(t) - \hat{\mathbf{r}}_{j}(t)} - \sum\limits_{j}\frac{1}{\hat{\mathbf{r}}_{i}(t-\Delta t) -\hat{\mathbf{r}}_{j}(t-1)},
\end{equation}
where $\hat{\mathbf{r}}_{j}$ are all particles in the rod.
The reward for the rod rotation itself is more involved as a partitioning scheme is used in order to compute a meaningful single-agent contribution which is applied to the total reward as:
\begin{equation}
    \centering
    r^{\text{rod rotation}}_{i} = \frac{\mathbf{\tau}_{i}}{\tau_{\text{net}}}\cdot\left(\omega_{\text{rod}}(t) - \omega_{\text{rod}}(t-\Delta t)\right),
\end{equation}
where $\tau_{i}$ is the torque exerted on the rod by colloid $i$, $\tau_{\text{net}}$ is the net torque acting on the rod, and $\omega_{\text{rod}}(t)$ is the angular velocity of the rod at time $t$.
The final reward for the colloid is then computed with the following:
\begin{equation}
    \centering
    r_{i} = \alpha\cdot r^{\text{rod proximity}}_{i} + \beta \cdot r^{\text{rod rotation}}_{i},
\end{equation}
where $\alpha$ and $\beta$ are scaling factors chosen in this work to be 10 and 100, respectively, such that successful rod rotation dominates the reward.
The results of the rod rotation experiment are outlined in Figures~\ref{fig:experiment-2} A)-C).
During the rod rotation experiments, two observables were of interest: the angular velocity of the rod, which measures the agents' efficacy, and the agents' distribution along the rod, which provides a description of their strategy.
Figure~\ref{fig:experiment-2}B) shows the average absolute rod velocity as a function of temperature.
In order to account for the natural motion of the rod, simulations have been performed without active colloids and these rod velocities subtracted from the active values.
It is clear from the measurements that the agents are capable of achieving greater rod velocities at higher temperatures.
This is of interest as the 0 K simulation provides the least resistance to rotation as no effects of the Brownian motion present.
As the temperature continues to increase, the mean rod velocity oscillates and decreases, suggesting that at some stage, the random fluctuations again become problematic for the agents' strategy.

The policy of the agents can be seen in Figures~\ref{fig:experiment-2} A) and C).
Figure~\ref{fig:experiment-2}A) plots the 2D histogram of the colloids for one of the 150 K simulation.
It is evident in this plot that the agents form on either end of the rod in a manner that maximises the applied torque.
A clearer picture of this strategy emerges when studying only the single-dimensional histogram along the long axis of the rod, shown in Figure~\ref{fig:experiment-2}C).
Here we can see that concentration along the ends of the rod is an emergent strategy for all temperatures.
This concentration is maximized at 150 K, possibly explaining the enhanced angular velocity in Figure~\ref{fig:experiment-2}C).
As temperature increases it seems that the agents move towards the center of the rod, perhaps preventing them for sliding off due to the additional random forces.
This trend is broken in the 350 K simulation where the agents once again form on the ends of the rod, although in a more asymmetric fashion, perhaps suggesting an alternative strategy in these conditions.

\section{Conclusion and Outlook}
\label{sec:conclusion-and-outlook}
We have performed experiments using the ESPResSo simulation engine to understand how temperature impacts the emergent strategy and efficacy of this strategies in micro-scale agents governed by MARL.
In the location detection experiments resembling biological chemotaxis, we found that increasing the temperature resulted in the agents taking direct approaches toward the source of the field. 
In contrast, with no Brownian motion, they entered a decaying orbit.
This was accompanied by the agents equilibrating closer to the source at higher temperatures, suggesting the strategy was more effective than orbiting.
The most exciting outcome was the policy adopted by the agents as we saw them take the translation action upon a move closer to the source and more rotation as they moved away.
This approach strategy is heavily reminiscent of bacterial run and tumble motion.
Rod rotation also demonstrated an evolving strategy with temperature as we saw the colloids favor sitting on the ends of the rods, thereby maximizing the applied torque. 
However, after 150 K the colloids began to favor moving inwards along the rod, presumably to avoid sliding off during rotation.
These results demonstrate that the environments of microscopic, intelligent agents profoundly impact their efficacy and the strategies they need to adopt to achieve different tasks.
Continued work in this area could look into more efficient training strategies to achieve the performance of the models at 150 K in the 350 K simulations.
Further, training models in simulation environments to be robust against these environmental factors could result in a smaller generalization gap for crossing into physical experiments.
Micro-robotics and MARL are, without a doubt, technologies of the future.
The key to unlocking their success lies in understanding the conditions under which they will work and using these to their advantage.

\section{Acknowledgements}
\label{sec:acknowledgements}
V.L.H and C.B acknowledge funding from the DFG Centre of Excellence 2117, Germany ''Centre for the
Advances Study of Collective Behaviour'', ID: 422037984.
C.H and S.T acknowledge financial support from the German Funding Agency (Deutsche Forschungsgemeinschaft DFG) under Germany’s Excellence Strategy EXC 2075-390740016, and S. T was supported by a LGF stipend of the state of Baden-W\"{u}rttemberg.
C.H, and S.T acknowledge financial support from the German Funding Agency (Deutsche Forschungsgemeinschaft DFG) under the Priority Program SPP 2363.

\bibliographystyle{plainnat}
\bibliography{main}

\begin{thebibliography}{28}
\providecommand{\natexlab}[1]{#1}
\providecommand{\url}[1]{\texttt{#1}}
\expandafter\ifx\csname urlstyle\endcsname\relax
  \providecommand{\doi}[1]{doi: #1}\else
  \providecommand{\doi}{doi: \begingroup \urlstyle{rm}\Url}\fi

\bibitem[B{\"a}uerle et~al.(2020)B{\"a}uerle, L{\"o}ffler, and
  Bechinger]{baeuerle20a}
Tobias B{\"a}uerle, Robert~C. L{\"o}ffler, and Clemens Bechinger.
\newblock Formation of stable and responsive collective states in suspensions
  of active colloids.
\newblock \emph{Nature Communications}, 11\penalty0 (1):\penalty0 2547, May
  2020.
\newblock ISSN 2041-1723.
\newblock \doi{10.1038/s41467-020-16161-4}.
\newblock URL \url{https://doi.org/10.1038/s41467-020-16161-4}.

\bibitem[Borra et~al.(2022)Borra, Biferale, Cencini, and Celani]{borra22a}
Francesco Borra, Luca Biferale, Massimo Cencini, and Antonio Celani.
\newblock Reinforcement learning for pursuit and evasion of microswimmers at
  low reynolds number.
\newblock \emph{Phys. Rev. Fluids}, 7:\penalty0 023103, Feb 2022.
\newblock \doi{10.1103/PhysRevFluids.7.023103}.
\newblock URL \url{https://link.aps.org/doi/10.1103/PhysRevFluids.7.023103}.

\bibitem[Bradbury et~al.(2018)Bradbury, Frostig, Hawkins, Johnson, Leary,
  Maclaurin, Necula, Paszke, Vander{P}las, Wanderman-{M}ilne, and
  Zhang]{bradbury18a}
James Bradbury, Roy Frostig, Peter Hawkins, Matthew~James Johnson, Chris Leary,
  Dougal Maclaurin, George Necula, Adam Paszke, Jake Vander{P}las, Skye
  Wanderman-{M}ilne, and Qiao Zhang.
\newblock {JAX}: composable transformations of {P}ython+{N}um{P}y programs,
  2018.
\newblock URL \url{http://github.com/google/jax}.

\bibitem[Brewer et~al.(2022)Brewer, Peltzer, and Lage]{brewer22a}
Peter~G. Brewer, Edward~T. Peltzer, and Kathryn Lage.
\newblock Life at low reynolds number re-visited: The efficiency of microbial
  propulsion.
\newblock \emph{Deep Sea Research Part I: Oceanographic Research Papers},
  185:\penalty0 103790, 2022.
\newblock ISSN 0967-0637.
\newblock \doi{https://doi.org/10.1016/j.dsr.2022.103790}.
\newblock URL
  \url{https://www.sciencedirect.com/science/article/pii/S0967063722001030}.

\bibitem[Chen and Bechinger(2022)]{chen22a}
Chun-Jen Chen and Clemens Bechinger.
\newblock Collective response of microrobotic swarms to external threats.
\newblock \emph{New Journal of Physics}, 24\penalty0 (3):\penalty0 033001, mar
  2022.
\newblock \doi{10.1088/1367-2630/ac5374}.
\newblock URL \url{https://dx.doi.org/10.1088/1367-2630/ac5374}.

\bibitem[Hartl et~al.(2021)Hartl, Hübl, Kahl, and Zöttl]{hartl21a}
Benedikt Hartl, Maximilian Hübl, Gerhard Kahl, and Andreas Zöttl.
\newblock Microswimmers learning chemotaxis with genetic algorithms.
\newblock \emph{Proceedings of the National Academy of Sciences}, 118\penalty0
  (19):\penalty0 e2019683118, 2021.
\newblock \doi{10.1073/pnas.2019683118}.
\newblock URL \url{https://www.pnas.org/doi/abs/10.1073/pnas.2019683118}.

\bibitem[Heek et~al.(2023)Heek, Levskaya, Oliver, Ritter, Rondepierre, Steiner,
  and van {Z}ee]{heek23a}
Jonathan Heek, Anselm Levskaya, Avital Oliver, Marvin Ritter, Bertrand
  Rondepierre, Andreas Steiner, and Marc van {Z}ee.
\newblock {F}lax: A neural network library and ecosystem for {JAX}, 2023.
\newblock URL \url{http://github.com/google/flax}.

\bibitem[Huijben et~al.(2022)Huijben, Kool, Paulus, and van Sloun]{huijben22a}
Iris A.~M. Huijben, Wouter Kool, Max~B. Paulus, and Ruud J.~G. van Sloun.
\newblock A review of the gumbel-max trick and its extensions for discrete
  stochasticity in machine learning, 2022.

\bibitem[Kingma and Ba(2017)]{kingma17a}
Diederik~P. Kingma and Jimmy Ba.
\newblock Adam: A method for stochastic optimization, 2017.

\bibitem[Krippendorf and Spannowsky(2022)]{krippendorf22a}
Sven Krippendorf and Michael Spannowsky.
\newblock A duality connecting neural network and cosmological dynamics.
\newblock \emph{Machine Learning: Science and Technology}, 3\penalty0
  (3):\penalty0 035011, aug 2022.
\newblock \doi{10.1088/2632-2153/ac87e9}.
\newblock URL \url{https://dx.doi.org/10.1088/2632-2153/ac87e9}.

\bibitem[Lavergne et~al.(2019)Lavergne, Wendehenne, Bäuerle, and
  Bechinger]{lavergne19a}
François~A. Lavergne, Hugo Wendehenne, Tobias Bäuerle, and Clemens Bechinger.
\newblock Group formation and cohesion of active particles with visual
  perception\&\#x2013;dependent motility.
\newblock \emph{Science}, 364\penalty0 (6435):\penalty0 70--74, 2019.
\newblock \doi{10.1126/science.aau5347}.
\newblock URL \url{https://www.science.org/doi/abs/10.1126/science.aau5347}.

\bibitem[Littman(1994)]{littman94a}
Michael~L. Littman.
\newblock Markov games as a framework for multi-agent reinforcement learning.
\newblock In William~W. Cohen and Haym Hirsh, editors, \emph{Machine Learning
  Proceedings 1994}, pages 157--163. Morgan Kaufmann, San Francisco (CA), 1994.
\newblock ISBN 978-1-55860-335-6.
\newblock \doi{https://doi.org/10.1016/B978-1-55860-335-6.50027-1}.
\newblock URL
  \url{https://www.sciencedirect.com/science/article/pii/B9781558603356500271}.

\bibitem[Lolli et~al.(2022)Lolli, Corsi, and DeSimone]{lolli22a}
Alberto Lolli, Giovanni Corsi, and Antonio DeSimone.
\newblock Control and navigation problems for model bio-inspired microswimmers.
\newblock \emph{Meccanica}, 57\penalty0 (10):\penalty0 2431--2445, Oct 2022.
\newblock ISSN 1572-9648.
\newblock \doi{10.1007/s11012-022-01567-9}.
\newblock URL \url{https://doi.org/10.1007/s11012-022-01567-9}.

\bibitem[Moreau et~al.(2021)Moreau, Ishimoto, Gaffney, and Walker]{moreau21a}
Clément Moreau, Kenta Ishimoto, Eamonn~A. Gaffney, and Benjamin~J. Walker.
\newblock Control and controllability of microswimmers by a shearing flow.
\newblock \emph{Royal Society Open Science}, 8\penalty0 (8):\penalty0 211141,
  2021.
\newblock \doi{10.1098/rsos.211141}.
\newblock URL
  \url{https://royalsocietypublishing.org/doi/abs/10.1098/rsos.211141}.

\bibitem[Muiños-Landin et~al.(2021)Muiños-Landin, Fischer, Holubec, and
  Cichos]{landin21a}
S.~Muiños-Landin, A.~Fischer, V.~Holubec, and F.~Cichos.
\newblock Reinforcement learning with artificial microswimmers.
\newblock \emph{Science Robotics}, 6\penalty0 (52):\penalty0 eabd9285, 2021.
\newblock \doi{10.1126/scirobotics.abd9285}.
\newblock URL
  \url{https://www.science.org/doi/abs/10.1126/scirobotics.abd9285}.

\bibitem[Qin et~al.(2023)Qin, Zou, Zhu, and Pak]{qin23a}
Ke~Qin, Zonghao Zou, Lailai Zhu, and On~Shun Pak.
\newblock {Reinforcement learning of a multi-link swimmer at low Reynolds
  numbers}.
\newblock \emph{Physics of Fluids}, 35\penalty0 (3), 03 2023.
\newblock ISSN 1070-6631.
\newblock \doi{10.1063/5.0140662}.
\newblock URL \url{https://doi.org/10.1063/5.0140662}.
\newblock 032003.

\bibitem[Schmidt et~al.(2020)Schmidt, Medina-S{\'a}nchez, Edmondson, and
  Schmidt]{schmidt20a}
Christine~K. Schmidt, Mariana Medina-S{\'a}nchez, Richard~J. Edmondson, and
  Oliver~G. Schmidt.
\newblock Engineering microrobots for targeted cancer therapies from a medical
  perspective.
\newblock \emph{Nature Communications}, 11\penalty0 (1):\penalty0 5618, Nov
  2020.
\newblock ISSN 2041-1723.
\newblock \doi{10.1038/s41467-020-19322-7}.
\newblock URL \url{https://doi.org/10.1038/s41467-020-19322-7}.

\bibitem[Soto et~al.(2020)Soto, Wang, Ahmed, and Demirci]{soto20a}
Fernando Soto, Jie Wang, Rajib Ahmed, and Utkan Demirci.
\newblock Medical {Micro/Nanorobots} in precision medicine.
\newblock \emph{Adv Sci (Weinh)}, 7\penalty0 (21):\penalty0 2002203, October
  2020.

\bibitem[Sugi et~al.(2021)Sugi, Ito, and H~Nagai]{takuma21a}
Takuma Sugi, Hiroshi Ito, and Ken H~Nagai.
\newblock Collective pattern formations of animals in active matter physics.
\newblock \emph{Biophys Physicobiol}, 18:\penalty0 254--262, October 2021.

\bibitem[Sutton and Barto(2018)]{sutton98a}
Richard~S. Sutton and Andrew~G. Barto.
\newblock \emph{Reinforcement Learning: An Introduction}.
\newblock The MIT Press, second edition, 2018.
\newblock URL \url{http://incompleteideas.net/book/the-book-2nd.html}.

\bibitem[van Ginkel et~al.(2021)van Ginkel, van Gisbergen, and
  Redig]{ginkel21a}
Bart van Ginkel, Bart van Gisbergen, and Frank Redig.
\newblock Run-and-tumble motion: The role of reversibility.
\newblock \emph{Journal of Statistical Physics}, 183\penalty0 (3):\penalty0 44,
  Jun 2021.
\newblock ISSN 1572-9613.
\newblock \doi{10.1007/s10955-021-02787-1}.
\newblock URL \url{https://doi.org/10.1007/s10955-021-02787-1}.

\bibitem[Walther and Müller(2013)]{walther13a}
Andreas Walther and Axel H.~E. Müller.
\newblock Janus particles: Synthesis, self-assembly, physical properties, and
  applications.
\newblock \emph{Chemical Reviews}, 113\penalty0 (7):\penalty0 5194--5261, 2013.
\newblock \doi{10.1021/cr300089t}.
\newblock URL \url{https://doi.org/10.1021/cr300089t}.
\newblock PMID: 23557169.

\bibitem[Weeks et~al.(1971)Weeks, Chandler, and Andersen]{weeks71a}
John~D Weeks, David Chandler, and Hans~C Andersen.
\newblock Role of repulsive forces in determining the equilibrium structure of
  simple liquids.
\newblock \emph{The Journal of chemical physics}, 54\penalty0 (12):\penalty0
  5237--5247, 1971.

\bibitem[Weik et~al.(2019)Weik, Weeber, Szuttor, Breitsprecher, de~Graaf,
  Kuron, Landsgesell, Menke, Sean, and Holm]{weik19a}
Florian Weik, Rudolf Weeber, Kai Szuttor, Konrad Breitsprecher, Joost de~Graaf,
  Michael Kuron, Jonas Landsgesell, Henri Menke, David Sean, and Christian
  Holm.
\newblock Espresso 4.0 -- an extensible software package for simulating soft
  matter systems.
\newblock \emph{The European Physical Journal Special Topics}, 227\penalty0
  (14):\penalty0 1789--1816, Mar 2019.
\newblock ISSN 1951-6401.
\newblock \doi{10.1140/epjst/e2019-800186-9}.
\newblock URL \url{https://doi.org/10.1140/epjst/e2019-800186-9}.

\bibitem[Zaverkin et~al.(2023)Zaverkin, Holzmüller, Bonfirraro, and
  Kästner]{zaverkin23a}
Viktor Zaverkin, David Holzmüller, Luca Bonfirraro, and Johannes Kästner.
\newblock Transfer learning for chemically accurate interatomic neural network
  potentials.
\newblock \emph{Phys. Chem. Chem. Phys.}, 25:\penalty0 5383--5396, 2023.
\newblock \doi{10.1039/D2CP05793J}.
\newblock URL \url{http://dx.doi.org/10.1039/D2CP05793J}.

\bibitem[Zhuang and Sitti(2016)]{zhuang16a}
Jiang Zhuang and Metin Sitti.
\newblock Chemotaxis of bio-hybrid multiple bacteria-driven microswimmers.
\newblock \emph{Scientific Reports}, 6\penalty0 (1):\penalty0 32135, Aug 2016.
\newblock ISSN 2045-2322.
\newblock \doi{10.1038/srep32135}.
\newblock URL \url{https://doi.org/10.1038/srep32135}.

\bibitem[Z\"{o}ttl and Stark(2023)]{zoettl}
Andreas Z\"{o}ttl and Holger Stark.
\newblock Modeling active colloids: From active brownian particles to
  hydrodynamic and chemical fields.
\newblock \emph{Annual Review of Condensed Matter Physics}, 14\penalty0
  (1):\penalty0 109--127, 2023.
\newblock \doi{10.1146/annurev-conmatphys-040821-115500}.
\newblock URL \url{https://doi.org/10.1146/annurev-conmatphys-040821-115500}.

\bibitem[Zöttl and Stark(2016)]{zoettl16a}
Andreas Zöttl and Holger Stark.
\newblock Emergent behavior in active colloids.
\newblock \emph{Journal of Physics: Condensed Matter}, 28\penalty0
  (25):\penalty0 253001, may 2016.
\newblock \doi{10.1088/0953-8984/28/25/253001}.
\newblock URL \url{https://dx.doi.org/10.1088/0953-8984/28/25/253001}.

\end{thebibliography}
\end{document}